\documentclass[%
 reprint,
 amsmath,amssymb,
 aps,
 pra,
]{revtex4-2}

\usepackage{graphicx}
\usepackage{dcolumn}
\usepackage{bm}
\usepackage{bbm}

\usepackage{hyperref}
\usepackage{breakurl}

\DeclareMathOperator{\tr}{tr}

\usepackage{xcolor}

\renewcommand\Re{\operatorname{Re}}
\renewcommand\Im{\operatorname{Im}}

\begin{document}

\preprint{APS/123-QED}

\title{Parametric approximation as open quantum systems problem}

\author{Artem Karasev}
\email{artem.karasev.01@gmail.com}

\affiliation{%
 Faculty of Physics, Lomonosov Moscow State University,
			Leninskie Gory, Moscow 119991, Russia
}%


\author{Alexander Teretenkov}
\email{taemsu@mail.ru}
\affiliation{
 Department of Mathematical Methods for Quantum Technologies, Steklov Mathematical Institute of Russian Academy of Sciences,
			ul. Gubkina 8, Moscow 119991, Russia
}%


\date{\today}

\begin{abstract}
In this work we develop an open quantum system view of the parametric approximation, which allows us to obtain systematic perturbative corrections to it. We consider the Jaynes-Cummings model with dissipation, assuming that the field is in the regime close to the parametric approximation with depletion. We obtain non-unitary corrections to the parametric approximation and additional dynamical Lamb-shift contributions to it. For high detuning, these non-unitary corrections appear to be non-Markovian before depletion. And we show that even after depletion, initial non-Markovian behaviour contributes to the dynamics via laser-induced polishing of the density matrix.
\end{abstract}

\maketitle


\section{\label{sec:introduction}Introduction
}

The parametric approximation, or constant amplitude approximation, is widely used for those modes of electromagnetic fields that have high intensity and has many applications \cite{Trovatello2021, Marty2020, Qiu2023, Allevi2006}. It can be regarded as a ''classical'' approximation of the quantum field \cite{Graham1984}. More precise conditions for the application of this approximation are discussed in~\cite{Paris1996, DAriano1999}. However, if one considers the parametric approximation as simply replacing the bosonic creation and annihilation operators by the fixed $c$-numbers, one completely neglects all quantum effects, including classical effects such as pump depletion. More importantly, there is no clear method for writing systematic corrections to the parametric approximation that take these effects into account.
		 
 One way is to move to {a} fully quantum regime and try to deal with the initial bosonic creation and annihilation operators without any reduction to $c$-numbers. However, using such an approach typically leads to problems that are not exactly solvable, even though they are solvable in the parametric approximation. Furthermore, if one solves such equations numerically by considering only the lowest Fock states, it becomes difficult to {handle} the high-intensity regime assumed by the parametric approximation. Another way is to assume that the bosonic mode is approximately in a coherent state \cite{Graham1984, Hillery1984, Accardi2003, Mabuchi2008}. In particular, this can be done in terms of the path-integral approach using the perturbative decomposition of the propagator of the system, which ultimately leads to perturbative corrections to the parametric approximation \cite{Hillery1984}. Alternatively, one can derive quantum stochastic equations in the Bogolubov-van Hove limit \cite{bogoliubov1946problems, van1955energy} and then obtain the master equations for the system density matrix in the field for which a parametric-like approximation is assumed \cite{Accardi2003}. It is also possible to obtain the Maxwell-Bloch-type equations \cite{Mabuchi2008}, i.e. the  coupled equations both for the two-level system density matrix and the parameter of the  coherent state.
	
We will consider the parametric approximation as an open quantum systems' problem where the reservoir is approximately in a coherent state. From an open quantum systems perspective, the path-integral approach to this problem is close to the Feynman-Vernon influence functional technique \cite[Section 3.6.4]{Breuer2002}, which can be used to solve harmonic systems \cite[Paragraph 3.6.4.4]{Breuer2002} and to derive non-perturbative numerical methods such as HEOM \cite{Tanimura1989}. The quantum stochastic equations approach is also widely used to describe both the system and the heat bath in the Markovian regime \cite{Accardi2002}, but its perturbative corrections require highly non-trivial multipole noises \cite{Pechen2002}. The Maxwell-Bloch-type equations \cite{Mabuchi2008} are non-linear in this case and hide their perturbative nature, even though in principle it can be revealed \cite{meretukov2024time}. For the systematic perturbative derivation of quantum master equations, however, the Nakajima-Zwanzig projection methods are usually used \cite[Section 9.2]{Breuer2002} in the time-convolutionless form \cite{Fulinski1967, Shibata1977}. In this case, the perturbative terms are provided by Kubo-van Kampen cumulants \cite{Kubo1963, VanKampen1974I, VanKampen1974II}. We will therefore follow this approach in our work.
	
We consider the simplest nontrivial system, the Jaynes-Cummings model. More precisely, to account for the irreversible depletion of the external electromagnetic field, we consider the dissipative Jaynes-Cummings model \cite{Sachdev1984}. If the initial states are chosen such that only a few lower Fock states of the bosonic mode are populated, then it can be solved explicitly. Moreover, the spectrum of such a model can be obtained exactly \cite{Torres2014}. However, if one is interested in the dynamics in cases where many Fock states are populated, then the approach in~\cite{Sachdev1984} leads to an infinite hierarchy of coupled equations, and ~\cite{Torres2014} requires finding eigenvectors of generic upper triangular matrices. Thus, even for this well-studied model, it is difficult to deal directly with the parametric regime. It is also worth noting that the methods of unitary perturbation theory, which take into account the reversible depletion of the pump \cite{Xing2023}, are not directly applicable in this case. However, in~\cite{Karasev2023}, it was shown that cumulant expansions for time-convolutionless master equations can be effectively applied to already open subsystems.
	
Note that in~\cite{Garraway1996, Garraway1997, Tamascelli2018, Tamascelli2019}, the equivalence conditions for the reduced dynamics of an open quantum system with two different types of environments are identified: one is a continuous bosonic environment leading to a unitary system-environment evolution, and the other is a discrete-mode bosonic environment leading to a system-mode (nonunitary) Lindbladian evolution with sums of zero-temperature dissipators. These effective discrete modes are called pseudomodes. Thus, the Jaynes-Cummings model with dissipation that we consider can be viewed not only as a result of the weak coupling derivation of the master equation for our system and some physical mode, but also as an effective master equation for our system and the pseudomode.

In Section \ref{sec:model}, we introduce the model and formulate the problem mathematically. In Section \ref{sec:TCLequation}, we describe our approach to its solution and obtain the first terms of the time-convolutionless master equation. We also discuss the physical effects these terms lead to. Section \ref{sec:polishing} is devoted to the contributions affecting the long-time behavior of the model.

\section{\label{sec:model}Jaynes-Cummings model with dissipation for approximately coherent mode}
Let us formulate the problem more precisely by writing down the master equation for the density matrix of the entire system \cite{Sachdev1984}.
\begin{equation}
\frac{d}{dt} \rho (t) = -i[H , \rho (t)] + \mathcal{D}\rho (t).
\end{equation}
The unitary part of the dynamics is defined by the Hamiltonian $H$, which decomposes into three parts. Subscripts $A$ and $B$ correspond to the system (two-level atom) and the bosonic mode, respectively.
\begin{eqnarray}
H =&& H_A + H_B + \lambda H_I, \nonumber \\ \nonumber 
H_A =&& \omega_A \sigma^+ \sigma^- {,} \\ 
H_B =&& \omega_c b^\dagger b .
\label{eq:freehamiltonian}
\end{eqnarray}
We claim that, for a given physical system, a difference $\omega_A-\omega_c = \Delta \omega$ is small. We factor out the dimensionless small parameter $\lambda$ from the coupling constant $g$, so that
\begin{equation}
H_I = g \sigma^+ b + \overline{g} \sigma^- b^\dagger.
\label{eq:interhamiltonian}
\end{equation}
The dissipative corrections arise due to the presence of the dissipator $\mathcal{D}$
\begin{equation}
\mathcal{D} = \gamma \left( b \cdot b^\dagger - \frac12 \{ b^\dagger b, \cdot \}  \right)
\label{eq:dissipator}
\end{equation}
and describe the Markovian decay of the bosonic mode.
  
In this work, we consider the interaction in the rotating wave approximation, so we use a local dissipator acting only {on the subsystem} $B$. Our perturbative decomposition can also be applied to interaction Hamiltonians beyond the rotating wave approximation, but in that case, we will need to consider other dissipators.
  
We use the projector
\begin{equation}
\mathcal{P} = \tr_B ( \; \cdot \; )\otimes | z \rangle \langle z|,
\label{eq:ourProjector}
\end{equation}
where $| z \rangle$ is a coherent state: $b | z \rangle = z | z \rangle$ and $\langle z | b^\dagger = \langle z | \overline{z}$. It has the Argyres-Kelley form \cite{Argyres1964} $\mathcal{P} = \tr_B ( \; \cdot \; )\otimes \rho_B$,
 where the Gibbs state is usually taken as $\rho_B$, but here we assume $\rho_B = | z \rangle \langle z|$ instead. The choice of the projector mathematically formalizes the main physical assumption that the external field is in an approximately coherent state from the point of view of our system.
  
We assume that the initial state of the system is consistent with the projector $\mathcal{P}\rho(0) = \rho (0)$. For the chosen projector, this means that the system and the reservoir are initially in the factorized state $\rho(0) = \rho_A(0) \otimes | z \rangle \langle z |$. The factorized form of \eqref{eq:ourProjector} also implies that this factorization approximately continues to hold during the evolution, analogous to the Born approximation in open quantum systems \cite[Section 3.3.1]{Breuer2002}.

\section{\label{sec:TCLequation}Time-convolutionless equations for systematic corrections to parametric approximation}

Let us extend the description of the master equation to a more general case of composite open systems \cite{aref2017holographic, saideh2020projection, Karasev2023, le2024adiabatic}, using the model discussed above as an example of such a composite open system. Let $\mathcal{L}_0$ be a generator corresponding to the free dynamics of the whole system. It is a superoperator, not necessarily consisting only of the unitary part, but we assume that it has the Gorini–Kossakowski–Sudarshan–Lindblad (GKSL) form. The superoperator $\mathcal{L}$ also satisfies the same condition, but it describes the interaction between the two subsystems.
\begin{equation}
\frac{d}{dt} \rho (t) = \left( \mathcal{L}_0 + \lambda\mathcal{L} \right) \rho (t),
\label{eq:basicDiffEq}
\end{equation}
Hence, $\mathcal{L}_0 = -i[H_A + H_B, \cdot \; ] + \mathcal{D}$ and $\mathcal{L} = -i[H_I, \cdot \; ] $. We denote superoperators by curly symbols $\mathcal{Y}$ and ordinary operators by straight $Y$. 
	
Further, we derive the equation for the reduced density matrix, so we deal with a system of ordinary differential equations, in contrast to the original infinite-dimensional situation. Thus, we will express the equation
\begin{equation}
\frac{d}{dt} \mathcal{P}\rho (t) = \sum_{n=0}^{N} \lambda^n \mathcal{K}_n \rho (t)
\end{equation}
up to any given order $N$. We call $\mathcal{K}_n$ the cumulant of order $n$. Such a derivation can be regarded as an adiabatic elimination \cite{finkelstein2020adiabatic, le2023heisenberg, riva2024explicit, tokieda2024complete, riva2024time, maity2024} of the degrees of freedom absent in the image of the projector $\mathcal{P}$.

Let us remark that our perturbative treatment is consistent with the fact that the RWA Hamiltonian, despite being exactly solvable without dissipation \cite{JaynesCummings1963, shore1993jaynes}, actually occurs by itself as a leading order of the algebraic perturbaton theory \cite{bogaevski2012algebraic, basharov2020global, basharov2024atom}. And small coupling assumed in the RWA Hamiltonian is consistent with local structure of dissipator \cite{trushechkin2021unified}. Let us also remark that the effective Hamiltonian \cite{trubilko2020hierarchy, basharov2021effective, trubilko2021effective} in general simplifies the calculation of the  dynamics of the vectorized creation and annihilation operators described in Appendix~\ref{app:vector}, so it seems that it should also be used if one applies our approach to more precise models than the Jaynes-Cummings model.
	
We provide the analytical derivation of the reduced equation up to and including the second order, since the second-order equation has a clear physical interpretation (see Appendix \ref{app:expansion}). The higher-order terms are calculated using mathematical software and are treated as refinements to the dynamics.

\subsection{\label{sec:freedynamics}Free dynamics of a two level system}
It is easy to see that the zero-order term is nothing but the free dynamics of the two-level system, as if there were no interaction with the environment (see Appendix \ref{app:freedyn}). Later, we will move to a rotating coordinate system in which this term will vanish.
\begin{equation}
\mathcal{K}_0 \rho = - i[\omega_A \sigma^+ \sigma^-, \rho_A] {.}
\end{equation}
Hereinafter, we omit the constant factor $\otimes | z \rangle\langle z |$ in all expressions for the cumulants.

In the following, we present computational results up to and including the fourth order. 

\subsection{\label{sec:firstorder}Parametric approximation with pump depletion}
Typically, the first term of perturbation theory vanishes when the reservoir is in an equilibrium state \cite[Section 9.2.2]{Breuer2002}, 
\begin{equation}
\mathcal{K}_1(t) = \mathcal{P} \mathcal{L}(t) \mathcal{P} = 0,
\end{equation}
since it has the meaning of the average of $\mathcal{L}(t)$ over the reservoir $\tr_B(\mathcal{L}(t) \rho_B) = 0$. However, in our case, this term will vanish only when $z = 0$.
Namely, $\mathcal{K}_1(t) \rho$ takes the following form (see Appendix \ref{app:parametric})
\begin{equation}
\mathcal{K}_1(t) \rho = -i e^{-\frac{\gamma}{2}t} [gz e^{-i \omega_c t} \sigma^+ + \overline{g} \overline{z} e^{i \omega_c t} \sigma^-, \rho_A ] .
\label{eq:Kumulant1}
\end{equation}
This term reproduces the parametric approximation for a two-level system interacting with a bosonic mode. However, it is corrected for energy losses associated with the dissipation of the electromagnetic field mode.
  
In our setup, the coherent state decays exponentially fast, so the contribution from $z$ to the master equations decreases rapidly with time. Nevertheless, at the initial stages of the system's evolution, this contribution plays a dominant role, especially if the parameter $\gamma$ is small.
  
\subsection{\label{sec:lambshift}Dynamical Lamb shift and non-Markovian effects}
In this section, we present the main result of the paper - an explicit form of the second-order term with time-dependent coefficients (see Appendix \ref{app:dissipation}). The second-order cumulant has a GKSL-like form and mathematically expresses two quantum effects. The first is the dynamical Lamb shift described by the commutator in (\ref{al:Kumulant2}). The second is written in the form of a dissipator, but in the space of a two-level system. Both of these coefficients would also arise when considering a conservative electromagnetic field, but in our case, they are corrected for losses in the subsystem $B$. 
\begin{eqnarray}
\mathcal{K}_2 (t) \rho = && \frac{-i |g|^2}{2 |\Gamma|^2}\Big( 2 \Delta \omega + \varphi_2 (t) \Big) [\sigma^+ \sigma^- , \rho_A] \nonumber \\ 
&&+\frac{|g|^2}{|\Gamma|^2} \Big( \gamma + f_2 (t) \Big) \left( \sigma^- \rho_A \sigma^+ - \frac12 \{ \sigma^+ \sigma^-, \rho_A \}\right),
\nonumber \\ 
\label{al:Kumulant2}
\end{eqnarray}
where $\Gamma = -i \Delta \omega + \frac{1}{2}\gamma$. The entire time dependence is contained in the functions $f_2 (t) = e^{-\frac{1}{2} \gamma t} \Big(2 \Delta \omega \sin \left(\Delta \omega t\right)- \gamma \cos \left(\Delta \omega t\right)\Big)$ and $\varphi_2 (t) = e^{-\frac{1}{2} \gamma t} \Big(-2\Delta \omega \cos \left( \Delta \omega t \right)- \gamma \sin \left( \Delta \omega t \right)\Big)$.

{The most direct effect of field depletion on the experimentally observable quantities is the change in the Mollow triplet. A ''naive'' approach would be to substitute the instantaneous depleted field in the usual formula for the Mollow spectrum for the undepleted field. However, such an approach completely ignores all quantum properties of the field. We can compare the results obtained with our quantum treatment (see Appendix~\ref{app:fluorescence} for details). Then we obtain a more squeezed but also more contrasted picture (in terms of the ratio of side peaks to intermediate minima) of the Mollow triplet, which is shown in Fig.~\ref{fig:Mollow}.

\begin{figure}
\includegraphics[width=1\linewidth]{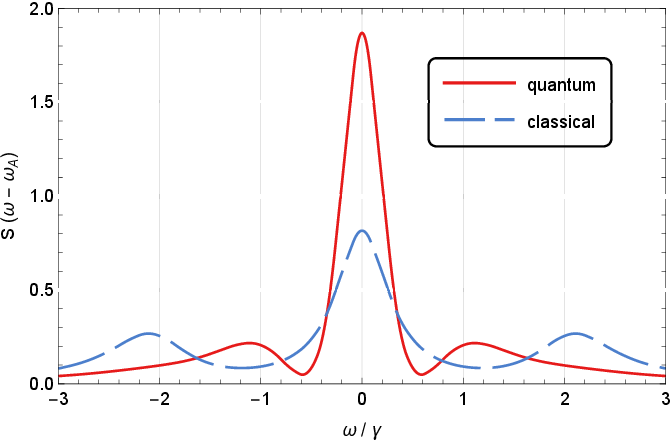}
\caption{\label{fig:Mollow} {The Mollow triplet based on classical and quantum treatment is shown for the specific choice of parameters
$\lambda = 0.4$, $g = \overline{g} = \gamma$, $z = \overline{z} =7 \gamma$, $\Delta \omega = 0 $, $\gamma t = 2$. 
The dashed blue line corresponds to the ''naive'' substitution of the depleted field into the formula for the undepleted case, ignoring all quantum effects. The solid red line corresponds to the quantum master equations obtained in this work. The ''naive'' treatment also has a delta function contribution in the center, which is omitted from the plot. }}
\end{figure}

}
  
The coefficients in the resulting expression contain oscillating functions, which is why, at small times, the resulting expression has only a GKSL-like form. Nevertheless, the functions $\varphi_2$ and $f_2$ decay at large times, which leads the system to Markov evolution. Thus, in the limit $t\to\infty$ the cumulant takes a GKSL form with Hamiltonian $H^{(2)} = ({\Delta \omega} |g|^2 / |\Gamma|^2) \sigma^+ \sigma^- $ and Lindblad operator $L^{(2)} = (\sqrt{\gamma} |g| / |\Gamma|) \sigma^- $.
  
Although the second-order equations are the most interesting and will be discussed further in Section \ref{sec:polishing}, our method also allows us to find expressions for the higher orders of perturbation theory. For the third and fourth orders, the calculations were carried out using Wolfram Mathematica \cite{mathematicaCode}. For this purpose, we introduced the method of vectorization of superoperators (see Appendix \ref{app:vector}). With this approach, we obtain equations in matrix form, which can be rewritten in the original algebraic form afterwards.
  
We can symbolically express $\mathcal{K}_3$ by introducing a time-dependent complex coefficient $a^{(3)} (t)$.
\begin{eqnarray}
\mathcal{K}_3 (t) \rho&=& |a^{(3)}|[( e^{i \psi} \sigma^+ - e^{- i \psi} \sigma^- ), \rho_A ] \nonumber \\ & &+ 2 |a^{(3)}| \left( e^{i \psi}\sigma^+ \sigma^- \rho_A \sigma^+ + e^{- i \psi} \sigma^- \rho_A \sigma^+ \sigma^- \right),
\nonumber \\
\label{eq:Kumulant3}
\end{eqnarray}
where
\begin{equation}
a^{(3)} (t) = g |g|^2 z e^{-i\omega_A t} \frac{e^{- \Gamma t} \left(1 - \Gamma t - e^{-\Gamma t} \right) }{\Gamma^2} \equiv |a^{(3)}|e^{i\psi}.
\end{equation}
~(\ref{eq:Kumulant3}) can be equivalently rewritten in GKSL-like (see~\cite{Hall2014}) form
\begin{eqnarray}
\mathcal{K}_3(t) &=& -i [H^{(3)}, \; \cdot \; ] \nonumber \\&&+ \sum_{j=1}^{2} (-1)^{j} \Big( L^{(3)}_j \; \cdot \;{L^{(3)}_j}^\dagger - \frac12 \{ {L^{(3)}_j}^\dagger L^{(3)}_j, \; \cdot \;\} \Big), 
\nonumber \\
\end{eqnarray}
where $H^{(3)}= |a^{(3)}|(i e^{i \psi} \sigma^+ - i e^{- i \psi} \sigma^- )$ and two Lindblad operators $L^{(3)}_1 = \sqrt{|a^{(3)}|}(\sigma^- - e^{i \psi} \sigma^+ \sigma^-)$ and $L^{(3)}_2 = \sqrt{|a^{(3)}|}(\sigma^- + e^{-i \psi} \sigma^+ \sigma^-)$.

{Remark that  the third-order contribution to the dissipator couples the populations and coherences, in contrast to~(\ref{al:Kumulant2}), which only leads to a decoupled classical Markovian dynamics for populations. In this sense, the second-order contribution captures only classical dissipation  effects, while the higher-order contributions essentially account for quantum dissipation effects. The higher-order terms also affect the instantaneous steady state of the dissipator, which is important for applications of master equations in quantum thermodynamics  \cite{dann2021quantum}.}
  
The form of~(\ref{eq:Kumulant3}) is significantly different from~(\ref{al:Kumulant2}), so the stationary states of the second- and third-order equations are different. The complex structure of~(\ref{eq:Kumulant3}) at the initial stages of evolution is also related to the effect we call the polishing of the density matrix.
This term tends to zero and completely disappears at large times $\mathcal{K}_3 (+\infty) = 0$. However, this is not the case for even orders. For example, the cumulant $\mathcal{K}_4$ at $t \to \infty$ does not go to zero. The non-vanishing part of the fourth-order cumulant $\mathcal{K}_4(+\infty)$ is a correction to the unitary and dissipative parts of the second-order generator with $H^{(4)} = |g|^4 \frac{\Delta \omega (\gamma^2 - |\Gamma|^2)}{|\Gamma|^6} \sigma^+ \sigma^-$ and Lindblad operator $L^{(4)} = |g|^2 \sqrt{\frac{\gamma ||\Gamma|^2 - 4 \Delta \omega^2|}{|\Gamma|^6} }\sigma^- $, {which define the Lindblad contribution for $|\Gamma|^2 - 4 \Delta \omega^2 \geq 0$ and the Lindblad-like contribution with a negative sing for $|\Gamma|^2 - 4 \Delta \omega^2 < 0$. Thus, this term lowers the long-time population and coherence decay rates in the significantly off-resonance case.} Remark that the role of the fourth-order time-convolutionless  equations in the usual open quantum systems setup is an actively discussed topic by itself \cite{breuer2006non, fruchtman2016perturbative, Crowder2024, suarez2024dynamics, xia2024markovian, chen2025benchmarking}.

\section{\label{sec:polishing}Laser-induced polishing of the density matrix}
  
From a physical point of view, we are interested in the dynamics of the system in the equilibrium regime. This regime arises after the decay of the correlations between the system and the reservoir and corresponds to the equations in the Markov approximation. We can switch to the required time scale by using the Bogolubov-van Hove rescaling. We rescale the time $t \to t/\lambda^2$ and apply the limit $\lambda \to 0$. Then we only need to have expressions for the cumulants up to second order. First, we need to solve the zero order equation and perform a coordinate substitution $\rho_A \to e^{- \mathcal{K}_0 t }\rho_A \equiv \tilde{\rho}$, this will allow us to use the procedure discussed above. The equation for times $\mathcal{O}(\lambda^{-2})$ takes a simple form. 
\begin{equation}
\frac{d}{dt} \tilde{\rho} = \mathcal{K}_2 (+\infty) \tilde{\rho} .
\label{eq:Olambda}
\end{equation}

{Qualitatively, we have the following picture. The interaction with the depleting field goes through two stages. In the first stage, the coherent interaction dominates. In the second stage, the field is almost completely depleted and the non-coherent interaction is dominant, but the effect of the first stage still contributes to the dynamics in the second stage by preparing the density matrix for the start of the second stage. We call this preparation "laser-induced polishing". In terms of the master equation at the first stage $\mathcal{K}_1 \neq 0$ but it tends exponentially fast to zero. And in the second state the density matrix satisfies~(\ref{eq:Olambda}), but the "laser-induced polishing" is manifested in the renormalization of the initial conditions. Such renormalization is expressed by the operator $\mathcal{R}$ (see Appendix \ref{app:matching}).}
\begin{equation}\label{eq:renormalization}
\mathcal{R}\rho(0) = \left( \mathcal{I} - i \lambda |\Gamma |^{-2} [gz \overline{\Gamma} \sigma^+ + \overline{g} \overline{z}  \Gamma \sigma^-, \;\cdot \;] \right) \rho(0) .
\end{equation}
From the experimenter's point of view, we observe an almost instantaneous excitation of $-\lambda\frac{2 }{|\Gamma|^2} \Re(i \overline{\Gamma} gz \tilde{\rho}_{01}(0))$ particles. Depending on the specific parameters of the model and the chosen initial state of the system, the matrix $\mathcal{R}\rho(0)$ may or may not be a density matrix. But, at least, $\tr\left(\mathcal{R}\rho(0)\right)=1$ and $\mathcal{R}\rho(0) = (\mathcal{R}\rho(0))^\dagger$.
  
In particular, the population of the excited state $\tilde{\rho}_{11}(t)$ of the two level system is still affected by initial coherence $\tilde{\rho}_{10}(0)$ for the long times
\begin{equation}
\tilde{\rho}_{11}(t) = e^{-\frac{|g|^2 \gamma t}{|\Gamma|^2}} \Big(\tilde{\rho}_{11}(0) - \lambda\frac{2 }{|\Gamma|^2} \Re(i \overline{\Gamma} gz \tilde{\rho}_{01}(0))\Big)
\label{eq:corrsolution}
\end{equation}
despite the fact that  dynamics of populations and coherences of the two level system are already decoupled for such long times.

\begin{figure}
\includegraphics[width=1\linewidth]{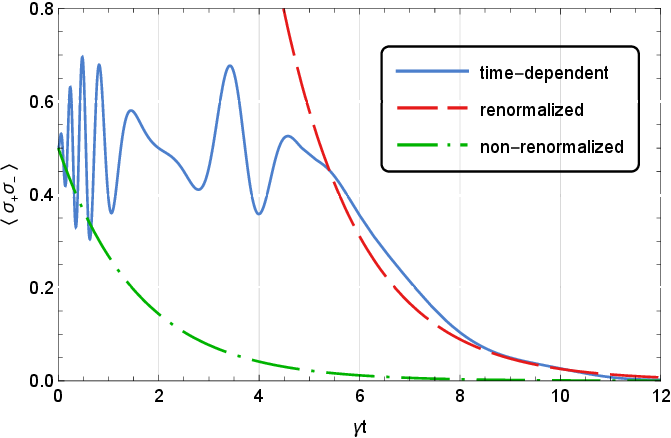}
\caption{\label{fig:intensity} {
All three graphs show the average $\langle \sigma_+ \sigma_- (t)\rangle $ as a function of $\gamma t$, i.e. of rescaled time in the units of the bosonic mode decay time, for a specific choice of parameters $\lambda = 0.055$, $g = \overline{g} = 20 \gamma$, $z = \overline{z} =342 \gamma$, $\Delta \omega   = 1.3 \gamma $. $\rho(0)$ is the eigenstate of $\sigma^x$ corresponding to the eigenvalue $+1$. The  solid  blue line is obtained by the time-dependent generator. The  dashed  red line is obtained by the long-time equation  \eqref{eq:Olambda} with renormalized initial condition \eqref{eq:renormalization}. And  dot-dashed green line  is obtained by the long-time equation \eqref{eq:Olambda} with the initial condition taken without renormalization. The initial oscillatory behavior of the solid blue line corresponds to the polishing of the density matrix during the first stage.  The decaying behavior of the solid  blue line corresponds, where it is close to  dashed  red line, corresponds to the second stage, at the beginning of which the density matrix is already polished.  However,  without taking into account the  polishing  via renormalization of the initial density matrix, the long-time equations cannot correctly reproduce dynamics at the second stage, which is reflected in the fact that the dot-dashed green line is far from solid  blue line during most of the dynamics.}}
\end{figure}

{We illustrate the physical picture discussed in this section in Fig.~\ref{fig:intensity}, where  $\langle \sigma_-\sigma_+ \rangle(t)$ as a function of time is presented, since it is proportional to the intensity of the radiated field \cite[Eq. (15.6.3)]{mandel1995optical} which  is experimentally measurable. Namely, we present the prediction of time-dependent second order master equations together with the long-time master equation~(\ref{eq:Olambda}) with and without renormalization.  A similar picture with a clear separation of two stages will occur under the condition that the decay rate in \eqref{eq:corrsolution} is smaller than the depletion rate in~\eqref{eq:Kumulant1}, but the field is strong enough to have a non-negligible contribution in \eqref{eq:renormalization}, i.e. if
\begin{equation}
|g| \lesssim |\Gamma|, \qquad \lambda |g| |z| \sim   |\Gamma|.
\end{equation}

Remark, that we choose the parameters such that coherent oscillations are clearly visible in Fig.~\ref{fig:intensity}, but for higher intensity of the initial input field they can be much faster, so the form from the experimental point of view they will look like stochastic behavior around some averaged dynamics, while the second stage has more regular behavior. In this case, the laser-induced polishing will look like preparation of the density matrix during some stochastic behavior for the more regular dynamics.
}

\section{\label{sec:quasiprobability}Dynamics for arbitrary initial state of bosonic mode}

Since we have obtained the solution of the second-order equation for the case of a coherent state, i.e. $\rho_B = | z \rangle\langle z |$, it is pretty straightforward to extend our calculations to cases where $\rho_B $ is expanded through Glauber–Sudarshan $P$ representation as $\rho_B = \int P(z)| z \rangle\langle z | d^2 z $. The use of such a reperesentation is dictated by the fact that we know how to solve equations in the diagonal representation. For instance, suppose we are interested in the situation when the external field is in the Schrödinger cat state  \cite{dodonov1974even} which can naturally arise  in the Jaynes-Cummings model without dissipation \cite{buvzek1992schrodinger}.
\begin{equation}
|\mathrm{cat} \rangle = c_\alpha \big(|\alpha\rangle+ e^{i\theta} |-\alpha\rangle\big)
\label{eq:catstate}
\end{equation}
The {whole} dependence on $z$ and $\overline{z}$ is in the {renormalization} superoperator $\mathcal{R}$.
As a result, the dynamics of a two-level system, weakly interacting with a cat state mode, will evolve according to
\begin{equation}
\tilde{\rho}^{cat}(t) = e^{\mathcal{K}_2 (+\infty)t} \mathcal{Z} \mathcal{R} \rho(0) .
\end{equation}
In this case the action of $\mathcal{Z} \equiv \int P(z) \cdot d^2 z$ will be in replacing 
\begin{equation}
z \to \alpha \frac{- i \sin(\theta) e^{-2|\alpha^2|}}{\left(1 + \cos(\theta)e^{-2|\alpha|^2}\right)}. 
\label{eq:cattransformation}
\end{equation}
Remark that if $\mathcal{Z}\mathcal{R} = \mathcal{I}$, then the laser polishing effect and the averaging with respect to the Glauber–Sudarshan quasiprobablity compensate each other. In our case for $\theta = 0$, i.e. for the even Schrödinger cat state, there is no change of the density matrix $\rho(0)$. Thus, long-time two-level system density matrix dynamics for such an exotic state of the bosonic mode is fully the same as for the vacuum state.

Let us also remark that renormalization of initial conditions be sometimes interpreted as manifestation of initial non-Markovian effects to long-time Markovian dynamics \cite{teretenkov2021non, teretenkov2021long, trushechkin2023long, trushechkin2024kinetic}, so an interesting question for the further study is  if such a compensation can restore the regression formulae for multi-time correlation functions as well.

This example also motivates to consider arbitrary quantizers   \cite{man2002alternative, andreev2020quantizer, man2020integral, adam2021properties} instead of $| z \rangle\langle z |$ in the projector $\mathcal{P}$ along with the correspondent operator symbols instead of the Glauber–Sudarshan quasiprobablity. {This generalization to other quantizers can also be used to extend our approach to other types of environmental degrees of freedom instead of the bosonic mode, e.g. to tensor products of qubits or oscillating spin-1/2 particles. It seems to be particularly effective for those states for which their probability representation \cite{man2021probability} is explicitly known together with the corresponding quantizers, e.g. for cat states of two or three qubits, or of an oscillating spin-1/2 particle \cite{mechler2024even}.}

\section{\label{sec:conclusions}Conclusions}
	
This work provides an open quantum systems view on the parametric approximation. In particular, this approach allows us to develop a systematic technique for obtaining perturbative corrections to the parametric approximation. Moreover, such a view gives us not only computational techniques but also reveals physical phenomena from open quantum systems theory, such as non-Markovian contributions to dynamics and initial state renormalization in the long-time Markovian regime for the parametric approximation.
	
We have applied this approach to the Jaynes-Cummings model with dissipation. We have reproduced the parametric approximation taking into account pump depletion in the first order of perturbation theory. In the second order of perturbation theory, the perturbation fails to be unitary and is described by a GKSL-like equation. The Lamb shift contribution exhibits time-dependent oscillatory behavior against the overall depletion. Additionally, the dissipator has a GKSL-like, but not a GKSL, form for high detuning. So it describes essentially non-Markovian behavior, which becomes Markovian only after depletion. Nevertheless, non-Markovian dynamics during the depletion still contributes to the initial condition for the Markovian long-time dynamics due to laser-induced polishing of the density matrix.

\begin{acknowledgments}
The authors thank Yu. M. Belousov, Kh. Sh. Meretukov, R. Singh and A. S. Trushechkin for the fruitful discussions of the problems considered in this work. {The authors thank the reviewers for their comments, which led to significant improvement of the paper.}
	
The work of A. Karasev was supported by the Foundation for the Advancement of Theoretical Physics and Mathematics “BASIS” under the grant № 24-2-1-48-1.
\end{acknowledgments}

\appendix

\section{Perturbative expansion}\label{app:expansion}
	
In a generic case, the real parts of all eigenvalues of a GKSL generator are non-positive, so the expression $\mathcal{P}e^{-\mathcal{L}_0 t}\rho$ may contain unboundedly increasing elements. For this reason, the transition to the interaction representation will be mathematically correct only if some conditions are fulfilled \cite{Karasev2023}. We solve this problem by considering the equations in the initial Schrödinger representation. This leads to better convergence of the final expressions, but also complicates the formulas. As a result, the inverse free dynamics $\mathcal{O}_0^{-1}$ appears in the terms of the expansion as well as it causes the presence of the zero term of perturbation theory.
\begin{widetext}
\begin{equation}
\mathcal{K}_n (t) = \sum_{q=0}^{n} (-1)^q \sum_{\sum_{j=0}^q k_j =n, k_j \geqslant 0} \dot{\mathcal{O}}_{k_0} \mathcal{O}^{-1}_0 \mathcal{O}_{k_1} \mathcal{O}^{-1}_0 ... \mathcal{O}^{-1}_0 \mathcal{O}_{k_q} \mathcal{O}^{-1}_0,
\label{eq:coeffKn}
\end{equation}
\end{widetext}
i.e. the inner sum runs over all possible compositions, 
\begin{eqnarray}
\mathcal{O}_k (t) &=& \int_{0}^t d t_1 \ldots \int_{0}^{t_{k-1}} d t_k \mathcal{P} e^{\mathcal{L}_0 t} \mathcal{L}(t_1) \ldots \mathcal{L}(t_k) \mathcal{P}, \nonumber \\ 
\mathcal{O}_0 (t) &=& \mathcal{P}e^{\mathcal{L}_0 t}\mathcal{P},
\label{eq:momentsDef}
\end{eqnarray}
where $\mathcal{L}(t) = e^{-\mathcal{L}_0 t} \mathcal{L} e^{\mathcal{L}_0 t}$.
	
We will call the superoperator $\mathcal{O}_n$ a moment of order $n$. We may write the first several cumulants explicitly in terms of these moments for better clarity
\begin{eqnarray}
\mathcal{K}_0(t) &=& \dot{\mathcal{O}}_0 \mathcal{O}^{-1}_0 ,\\ 
\mathcal{K}_1(t) &=& \dot{\mathcal{O}}_1 \mathcal{O}^{-1}_0 - \dot{\mathcal{O}}_0 \mathcal{O}^{-1}_0 \mathcal{O}_1 \mathcal{O}^{-1}_0, \\ 
\mathcal{K}_2(t) &=& \dot{\mathcal{O}}_2 \mathcal{O}^{-1}_0 - \dot{\mathcal{O}}_1 \mathcal{O}^{-1}_0 \mathcal{O}_1 \mathcal{O}^{-1}_0 - \dot{\mathcal{O}}_0 \mathcal{O}^{-1}_0 \mathcal{O}_2 \mathcal{O}^{-1}_0 \nonumber \\ 
&&+ \dot{\mathcal{O}}_0 \mathcal{O}^{-1}_0 \mathcal{O}_1 \mathcal{O}^{-1}_0 \mathcal{O}_1 \mathcal{O}^{-1}_0,
\label{eq:Kdef}
\end{eqnarray}
where $\mathcal{O}_1 = \mathcal{P}[e^{\mathcal{L}_0 t}, \mathcal{X}^{(1)}] \mathcal{P}$ and $\mathcal{X}^{(1)}$ is the solution to the equation $[\mathcal{L}_0 , \mathcal{X}^{(1)}]= \mathcal{L}$ and $\mathcal{O}_2 = \mathcal{P} [e^{\mathcal{L}_0 t}, \mathcal{X}^{(1)}] \mathcal{X}^{(1)}\mathcal{P} - \mathcal{P}[e^{\mathcal{L}_0 t}, \mathcal{X}^{(2)}] \mathcal{P}$ with $\mathcal{X}^{(2)}: [\mathcal{L}_0 , \mathcal{X}^{(2)}]= \mathcal{L}\mathcal{X}^{(1)}$.
 
\subsection{Free dynamics}\label{app:freedyn}
Let us write down the explicit form for generators (see Eqs.~ (\ref{eq:freehamiltonian}),(\ref{eq:interhamiltonian}) and (\ref{eq:dissipator}))
\begin{eqnarray}
\mathcal{L}_0 &=& -i [\omega_A \sigma^+ \sigma^- + \omega_c b^\dagger b, \; \cdot \;] \nonumber \\ &&+ \gamma \left( b \; \cdot \;b^\dagger- \frac12  b^\dagger b \; \cdot \; - \frac12 \; \cdot \; b^\dagger b \right), 
\label{eq:freeGen} \\ 
\mathcal{L} &=& -i [(g \sigma^+ b + \overline{g}\sigma^- b^\dagger ), \; \cdot \;]. 
\label{eq:intGen}
\end{eqnarray}
The free dynamics generator is easily factorized $\mathcal{L}_0 = \mathcal{L}^A_0 \otimes \mathcal{I}_B + \mathcal{I}_A \otimes \mathcal{L}^B_0$. 
\begin{eqnarray}
\mathcal{L}^A_0 &=& -i [\omega_A \sigma^+ \sigma^-, \; \cdot \;] ,\\ 
\mathcal{L}^B_0 &=& -i [\omega_c b^\dagger b, \; \cdot \;] + \gamma \left( b \; \cdot \;b^\dagger- \frac12  b^\dagger b \; \cdot \; - \frac12 \; \cdot \; b^\dagger b \right). \nonumber
\\
\end{eqnarray}
Both generators $\mathcal{L}^A_0$ and $\mathcal{L}^B_0$ have the GKSL form. Note that the coherent state remains coherent during free evolution \cite[Section 4.4.1]{Breuer2002}
\begin{equation}
e^{\mathcal{L}^B_0 t} | z \rangle \langle z | = | z(t) \rangle \langle z(t) |,
\end{equation}
where $z(t) = z e^{(-i\omega_c - \frac{\gamma}{2}) t }$.
	
To calculate $\mathcal{K}_0$ it is enough to know how $\mathcal{O}_0,\;\mathcal{O}^{-1}_0$ and $\dot{\mathcal{O}}_0$ act on the factorized density matrix at the coherent state of the field. 
\begin{equation}
\mathcal{O}_0 \rho = e^{\mathcal{L}^A_0 t} \rho_A \tr_B \left( e^{\mathcal{L}^B_0 t} | z \rangle\langle z | \right) \otimes | z \rangle\langle z |.
\end{equation}
Dynamical map $e^{\mathcal{L}^B_0 t} $ is trace-preserving, so $\tr_B \left( e^{\mathcal{L}^B_0 t} | z \rangle\langle z | \right) = \tr_B \left( | z (t) \rangle\langle z (t) | \right) = 1$.
\begin{equation}
\mathcal{O}^{-1}_0 (\rho_A \otimes | z \rangle \langle z | )= e^{-\mathcal{L}^A_0 t} \rho_A \otimes | z \rangle\langle z |.
\end{equation}
\begin{equation}
\dot{\mathcal{O}}_0 = \mathcal{L}^A_0 e^{\mathcal{L}^A_0 t} \cdot \tr_B \left( e^{\mathcal{L}^B_0 t} \; \cdot \;\right) + 
e^{\mathcal{L}^A_0 t} \cdot \tr_B \left( \mathcal{L}^B_0 e^{\mathcal{L}^B_0 t}\; \cdot \;\right).
\end{equation}
The only thing left to do is to use the trace preservation property
\begin{equation}
\tr_B \left( \mathcal{L}^B_0 e^{\mathcal{L}^B_0 t} | z \rangle \langle z | \right) = \tr_B \left( \mathcal{L}^B_0 | z (t) \rangle \langle z (t) | \right) = 0
\end{equation}
to obtain the final expression.
	
\subsection{Pseudoinverse superoperators}\label{app:pseudoinverse}
We aim to write down the final equations in algebraic form by integrating all expressions inside the moments. This will simplify the analysis of the master equation at large times. Consider the integral  $\int_{0}^t d t_1 \mathcal{L}(t_1) = [\mathcal{L}_0, \; \cdot \;]^{-1} (e^{- [\mathcal{L}_0, \; \cdot \;] t} \mathcal{L} - \mathcal{L}) = (e^{- [\mathcal{L}_0, \; \cdot \;] t} - I) [\mathcal{L}_0, \; \cdot \;]^{-1} \mathcal{L} $. Now it is clear why we expressed the moment $\mathcal{O}_1$ using $\mathcal{X}^{(1)} \equiv [\mathcal{L}_0, \; \cdot \; ]^{ -1 } \mathcal{L}$. The calculation of such pseudo-inverse commutation superoperators is discussed in~\cite{Nikolaev2016}, but in the situation of unitary dynamics.
	
To write the equation up to and including the second order, it is convenient to find the explicit form of $\mathcal{X}^{(1)}$. We can construct a closed algebra of superoperators with respect to the commutator action. After that we can easily reverse the commutator action and find the desired form of $\mathcal{X}^{(1)}$.

The superoperator $\mathcal{X}^{(1)} \equiv [\mathcal{L}_0, \; \cdot \; ]^{-1} \mathcal{L}$ may be written in the following form
\begin{eqnarray}
\mathcal{X}^{(1)} &=& \frac{-i g }{\Gamma}(\sigma^+ b \; \cdot\;) +
\frac{-i \overline{g}}{- \Gamma} \left( ( \sigma^- b^\dagger \; \cdot\;) - \frac{\gamma}{\overline{\Gamma}} (\sigma^- \cdot \;b^\dagger) \right) 
\nonumber \\
&&+ \frac{i g}{- \overline{\Gamma}} \left( (\;\cdot \; \sigma^+ b) - \frac{\gamma}{\Gamma} (b \; \cdot \;\sigma^+) \right) + \frac{i \overline{g}}{\overline{\Gamma}} (\;\cdot \; \sigma^- b^\dagger),
\label{eq:X1}
\nonumber \\
\end{eqnarray}
where $\Gamma = -i (\omega_A - \omega_c) + \frac12 \gamma$.

Let us find the result of action of the commutation superoperator $[\mathcal{L}_0, \cdot \;]$ on all the combinations of the creation and annihilation operators which preserve the number of particles. Then, it will be easy to find all necessary combinations and coefficients.
\begin{equation}
[\mathcal{L}_0 , (\sigma^- b^{\dagger} \cdot\;) ] = \left( i(\omega_A - \omega_c) - \frac12 \gamma \right) (\sigma^- b^{\dagger} \cdot\;) + \gamma (\sigma^- \; \cdot \; b^\dagger ). 
\end{equation}
Quite similarly

\begin{eqnarray}
[\mathcal{L}_0 , (\;\cdot \; \sigma^- b^{\dagger} ) ] &=& \overline{\Gamma} (\;\cdot \; \sigma^- b^{\dagger} ) \\
\left[ \mathcal{L}_0 , (\sigma^+ b \;\cdot\;)\right] &=& \Gamma (\sigma^+ b \; \cdot \;) \\
\left[ \mathcal{L}_0 , (\;\cdot \; \sigma^+ b ) \right] &=& -\overline{\Gamma} (\;\cdot \; \sigma^+ b) + \gamma ( b\cdot \sigma^+). 
\end{eqnarray}

All four combinations in the left-hand side of the equation are components of $\mathcal{L}$. The commutation added two more “non-diagonal”
\begin{eqnarray}
[\mathcal{L}_0 , (\sigma^- \cdot \; b^\dagger) ] &=& \overline{\Gamma} (\sigma^- \cdot \; b^{\dagger} )
\\
\left[\mathcal{L}_0 , (b \; \cdot \;\sigma^+)\right] &=& \Gamma (b \; \cdot \;\sigma^+).
\end{eqnarray}

Thus we have found the closed algebra of superoperators, it remains only to invert the coefficients.

\subsection{Parametric approximation}\label{app:parametric}
Starting from this stage we will use the explicit form of $\mathcal{X}^{(1)}$, even though we could do without it in the first order. For consistency, we will calculate the cumulants after integration of all expressions for the moments. We number all $6$ elements of $\mathcal{X}^{(1)}$ in the same order as they are in the sum~(\ref{eq:X1}). It is convenient for us to work with the elements of $\mathcal{X}^{(1)}$ separately.
\begin{equation}
\mathcal{X}^{(1)} = \sum_{i=1}^6 \mathcal{X}^{(1)}_i = \sum_{i=1}^6 \mathcal{X}^{(1)}_{A , i} \otimes \mathcal{X}^{(1)}_{B , i} 
\end{equation}
Then, expressions for $\mathcal{O}_1$ also decompose into summands.
\begin{eqnarray}
\dot{\mathcal{O}}_1 &=& \frac{d}{dt}\mathcal{P}[e^{\mathcal{L}_0 t}, \mathcal{X}^{(1)}] \mathcal{P} \nonumber\\ 
&=& \sum_{i} \Big( \mathcal{L}^A_0 e^{\mathcal{L}^A_0 t} \mathcal{X}^{(1)}_{A , i} \; \cdot \;\tr_B \left( e^{\mathcal{L}^B_0 t} \mathcal{X}^{(1)}_{B , i} \; \cdot \; \right) 
\nonumber \\ 
&&+ e^{\mathcal{L}^A_0 t} \mathcal{X}^{(1)}_{A , i} \; \cdot \;\tr_B \left(\mathcal{L}^B_0	e^{\mathcal{L}^B_0 t} \mathcal{X}^{(1)}_{B , i} \; \cdot \;\right)  \nonumber \\ 
&&- \mathcal{X}^{(1)}_{A , i} \mathcal{L}^A_0 e^{\mathcal{L}^A_0 t} \; \cdot \;\tr_B \left( \mathcal{X}^{(1)}_{B , i}	e^{\mathcal{L}^B_0 t} \; \cdot \; \right) 
\nonumber \\ 
&&- \mathcal{X}^{(1)}_{A , i} e^{\mathcal{L}^A_0 t} \; \cdot \;\tr_B \left(\mathcal{X}^{(1)}_{B , i} \mathcal{L}^B_0	e^{\mathcal{L}^B_0 t} \; \cdot \; \right) \Big) \otimes | z \rangle \langle z|. \nonumber
\\
\end{eqnarray}
In practice, the expressions are simplified due to the fact that $e^{\mathcal{L}^B_0 t}$ is a trace preserving dynamical map, so $\tr_B \left( e^{\mathcal{L}^B_0 t} \; \cdot \;\right) = \tr_B \left( \; \cdot \; \right)$ and $\tr_B \left( \mathcal{L}^B_0 \; \cdot \; \right) = 0 $. Besides, we perform calculations of such traces using the transition to Heisenberg representation $\tr_B \left( \mathcal{X}^{(1)}_{B , i} e^{\mathcal{L}^B_0 t} \; \cdot \;\right) = \tr_B \left( e^{(\mathcal{L}^B_0)^\dagger t} \mathcal{X}^{(1)}_{B , i}  \; \cdot \;\right)$.
\begin{eqnarray}
\dot{\mathcal{O}}_1 \mathcal{O}^{-1}_0 \rho &=& 
\sum_{i} \mathcal{L}^A_0 e^{\mathcal{L}^A_0 t} \mathcal{X}^{(1)}_{A , i} e^{-\mathcal{L}^A_0 t} \rho_A \tr_B \left( \mathcal{X}^{(1)}_{B , i} | z \rangle \langle z | \right) \nonumber \\
&&- \mathcal{X}^{(1)}_{A , i} \mathcal{L}^A_0 \rho_A \tr_B\left( e^{(\mathcal{L}^B_0)^\dagger t} \mathcal{X}^{(1)}_{B , i}	| z \rangle \langle z | \right) \nonumber \\ 
&&- \mathcal{X}^{(1)}_{A , i} \rho_A \tr_B \left( ( \mathcal{L}^B_0)^\dagger e^{(\mathcal{L}^B_0)^\dagger t} \mathcal{X}^{(1)}_{B , i} | z \rangle\langle z | \right).
\nonumber \\
\end{eqnarray}
Here is an example of a typical calculation for combination $\mathcal{X}^{(1)}_1 = (\sigma^+ b \; \cdot\;)$. Since $e^{\mathcal{L}^A_0 t} \sigma^+ e^{-\mathcal{L}^A_0 t} \rho_A = e^{- i\omega_A t} \sigma^+ \rho_A$ and $\tr_B\left( ( e^{ (\mathcal{L}^B_0)^\dagger t} b ) | z \rangle\langle z | \right) = e^{(-i\omega_c - \frac12 \gamma )t} z $, one can verify that
\begin{eqnarray}
\left( \dot{\mathcal{O}}_1 \mathcal{O}^{-1}_0 \right)_1 \rho &=& \Big( e^{- i\omega_A t} \mathcal{L}^A_0 (\sigma^+ \rho_A) 
\nonumber \\ 
&&-e^{(-i\omega_c - \frac12 \gamma )t} \sigma^+ \mathcal{L}^A_0 \rho_A
\nonumber \\
&&-(-i \omega_c - \frac12 \gamma )e^{(-i\omega_c - \frac12 \gamma )t} \sigma^+ \rho_A \Big)z,
\nonumber\\
\end{eqnarray}
\begin{equation}
\left(\dot{\mathcal{O}}_0 \mathcal{O}^{-1}_0 \mathcal{O}_1 \mathcal{O}^{-1}_0 \right)_1 \rho = ( e^{- i\omega_A t} - e^{(-i\omega_c - \frac12 \gamma )t} ) z \mathcal{L}^A_0 (\sigma^+ \rho_A) ,
\end{equation}
\begin{equation}
\left(\dot{\mathcal{O}}_1 \mathcal{O}^{-1}_0 - \dot{\mathcal{O}}_0 \mathcal{O}^{-1}_0 \mathcal{O}_1 \mathcal{O}^{-1}_0 \right)_1 \rho = - \overline{\Gamma} e^{(-i\omega_c - \frac12 \gamma )t} z \sigma^+ \rho_A.
\end{equation}
The index $1$ denotes the part of an expression associated with $\mathcal{X}^{(1)}_1$. Summing up similar expressions for all indices we obtain~(\ref{eq:Kumulant1}).
  
\subsection{Dissipation}\label{app:dissipation}
Second-order computation requires consideration of all possible quadratic expressions involving two-level operators that take into account the dominant side. Let us focus on $(\sigma^- \cdot \; \sigma^+)$ combination. The derivations for the other combinations are completely analogous.
 
Since $\mathcal{L}^A_0 (\sigma^- \cdot \; \sigma^+) = 0$, expressions $\dot{\mathcal{O}}_0 \dots = 0 $ for this particular combination.
  
We formally divide $\mathcal{O}_2$ into two independent parts (although, of course, $\mathcal{X}^{(2)}$ depends on $\mathcal{X}^{(1)}$): $\mathcal{O}_2 = (\mathcal{O}_2)_{\mathcal{X}^{(2)} = 0} + (\mathcal{O}_2)_{\mathcal{X}^{(1)} = 0}$.
\begin{enumerate}
\item Let $\mathcal{X}^{(2)} = 0$.
\begin{eqnarray}
&&\left( \dot{\mathcal{O}}_2 \mathcal{O}^{-1}_0 \right)_{\mathcal{X}^{(2)} = 0} =
\nonumber \\
&&\sum_{i \neq j}\mathcal{L}^A_0 e^{ \mathcal{L}^A_0 t} \mathcal{X}^{(1)}_{A , i} \mathcal{X}^{(1)}_{A , j} e^{- \mathcal{L}^A_0 t} \; \cdot \;\tr_B \left(\mathcal{X}^{(1)}_{B , i} \mathcal{X}^{(1)}_{B , j} \; \cdot\;\right) 
\nonumber \\ 
&&- \mathcal{X}^{(1)}_{A , i} \mathcal{L}^A_0 e^{ \mathcal{L}^A_0 t} \mathcal{X}^{(1)}_{A , j} e^{- \mathcal{L}^A_0 t} \; \cdot \;\tr_B \left(\mathcal{X}_{B , i}^{(1)} e^{ \mathcal{L}^B_0 t} \mathcal{X}_{B , j}^{(1)} \; \cdot \; \right) 
\nonumber \\
&&- \mathcal{X}_{A , i}^{(1)} e^{ \mathcal{L}^A_0 t} \mathcal{X}_{A , i}^{(1)} e^{- \mathcal{L}^A_0 t} \; \cdot \;\tr_B \left(\mathcal{X}^{(1)}_{B , i} \mathcal{L}^B_0 e^{ \mathcal{L}^B_0 t} \mathcal{X}^{(1)}_{B , j} \; \cdot \; \right)
\nonumber \\
\end{eqnarray}
For $\mathcal{X}^{(1)}_4 = \frac{-ig}{\overline{\Gamma}}(\; \cdot \; \sigma^+ b)$ and $\mathcal{X}^{(1)}_2 = \frac{i \overline{g}}{\Gamma}(\sigma^- b^\dagger \; \cdot \;)$ it leads to
\begin{equation}
\left( \dot{\mathcal{O}}_2 \mathcal{O}^{-1}_0 \right)_{\mathcal{X}^{(2)} = 0} \rho = \overline{\Gamma}\frac{g \overline{g}}{\Gamma \overline{\Gamma}}e^{-\overline{\Gamma} t}(1+z\overline{z})\sigma^- \rho_A \sigma^+ ,
\end{equation}
\begin{equation}
\dot{\mathcal{O}}_1 \mathcal{O}^{-1}_0 \mathcal{O}_1 \mathcal{O}^{-1}_0 \rho = \overline{\Gamma} \frac{g \overline{g}}{\Gamma \overline{\Gamma}}\left(e^{-\overline{\Gamma} t} - e^{-\gamma t}\right) z\overline{z} \sigma^- \rho_A \sigma^+ .
\end{equation}
Summing over $i \neq j$, we get the resulting expression for the combination $(\sigma^- \cdot \;\sigma^+)$:
\begin{eqnarray}
\left( \dot{\mathcal{O}}_2 \mathcal{O}^{-1}_0 \right)_{\mathcal{X}^{(2)} = 0} \rho - \dot{\mathcal{O}}_1 \mathcal{O}^{-1}_0 \mathcal{O}_1 \mathcal{O}^{-1}_0 \rho = 
\nonumber \\
\frac{g \overline{g}}{|\Gamma|^2} \Big(z\overline{z} \gamma e^{-\gamma t} + f_2(t) \Big) \sigma^- \rho_A \sigma^+ 
\end{eqnarray}

\item Consider the remaining part of $\mathcal{O}_2$, formally denoted as $(\mathcal{O}_2)_{\mathcal{X}^{(1)} = 0}$.
The part of interest of $\mathcal{L} \mathcal{X}^{(1)}$, which contains $(\sigma^- \cdot \;\sigma^+)$:
\begin{eqnarray}
&&\left(\mathcal{L} \mathcal{X}^{(1)} \right)_{(\sigma^- \cdot \;\sigma^+)} =
\nonumber \\
&&\frac{\gamma |g|^2}{|\Gamma|^2} (\sigma^- \cdot \sigma^+) \otimes \left( b^\dagger b \cdot + \cdot b^\dagger b - b^\dagger \cdot b \right) \equiv
\nonumber \\ &&\frac{\gamma |g|^2}{|\Gamma|^2} (\sigma^- \cdot \sigma^+) \otimes \mathcal{B}_3.
\end{eqnarray}
Using the rule of differentiation of an integral with a variable upper limit, we can compute $(\mathcal{O}_2)_{\mathcal{X}^{(1)} = 0}$ without explicitly knowing $\mathcal{X}_2$. More in detail, consider
\begin{eqnarray}
&&\left( \dot{\mathcal{O}}_2 \mathcal{O}_0^{-1} \right)_{(\sigma^- \cdot \;\sigma^+)}=
\nonumber \\
&& - \frac{d}{dt} \left( \mathcal{P} 	e^{\mathcal{L}_0 t } \int_{0}^{t} e^{-\mathcal{L}_0 t_1 } \left( \mathcal{L} \mathcal{X}^{(1)} \right)_{(\sigma^- \cdot \;\sigma^+)} e^{\mathcal{L}_0 t_1 } dt_1 \mathcal{P} \right)= 
\nonumber\\ 
&&- \frac{\gamma |g|^2}{|\Gamma|^2} (\sigma^- \cdot \; \sigma^+) \frac{d}{dt} \left( \tr_B \left( e^{\mathcal{L}^B_0 t } \int_{0}^{t} dt_1 e^{-[\mathcal{L}_0 , \; \cdot \; ] t } \mathcal{B}_3 \right) \right) =
\nonumber \\ 
&& - \frac{\gamma |g|^2}{|\Gamma|^2} (\sigma^- \cdot \; \sigma^+) \tr_B \left( \mathcal{B}_3 e^{\mathcal{L}^B_0 t} \; \cdot \;\right) = 
\nonumber \\
&&- \frac{\gamma |g|^2}{|\Gamma|^2} (\sigma^- \cdot \; \sigma^+) \tr_B \left( e^{(\mathcal{L}^B_0)^\dagger t}\mathcal{B}_3 \; \cdot \;\right).
\end{eqnarray}
So that the contribution arising from $\mathcal{X}^{(2)}$
\begin{equation}
\left( \dot{\mathcal{O}}_2 \mathcal{O}^{-1}_0 \right)_{\mathcal{X}^{(1)} = 0} \rho = \frac{g \overline{g}}{|\Gamma|^2} \Big(\gamma - \gamma e^{-\gamma t} z\overline{z} \Big) \sigma^- \rho_A \sigma^+. 
\end{equation}
\end{enumerate}
As a result,
\begin{equation}
\Big(\mathcal{K}_2 (t) \Big)_{(\sigma^- \cdot \;\sigma^+)} \rho = \frac{|g|^2}{|\Gamma|^2} \Big( \gamma + f_2 (t) \Big) \sigma^- \rho_A \sigma^+,
\end{equation}
which is a part of (\ref{al:Kumulant2}).

\section{Asymptotic matching}\label{app:matching}
Consider a finite-dimensional second-order equation
\begin{equation}
\frac{d}{dt} \rho_A = \mathcal{K}_0 \rho_A + \lambda \mathcal{K}_1 (t) \rho_A + \lambda^2 \mathcal{K}_2 (t) \rho_A.
\end{equation}
We will replace the variable in this way $\rho_A \to e^{- \mathcal{K}_0 t }\rho_A = \tilde{\rho}$ so that the term at $\lambda^0$ is reduced. Note that $\tilde{\rho}(0) = \rho_A(0)$.
\begin{eqnarray}\label{eq:NoZero}
\frac{d}{dt} \tilde{\rho} &=& e^{- \mathcal{K}_0 t } \left( \lambda \mathcal{K}_1 (e^{\mathcal{K}_0 t }\tilde{\rho}) + \lambda^2 \mathcal{K}_2 (e^{\mathcal{K}_0 t }\tilde{\rho}) \right) = 
\nonumber \\
&&\lambda \mathcal{K}^I_1 \tilde{\rho} + \lambda^2 \mathcal{K}^I_2 \tilde{\rho},
\end{eqnarray}
where $\mathcal{K}^I_i = e^{-[\mathcal{K}_0, \; \cdot \; ]t} \mathcal{K}_i$.
It is easy to verify that
\begin{eqnarray}
\mathcal{K}^I_1 &=& -i [gz e^{-\Gamma t} \sigma^+ + \overline{g} \overline{z} e^{- \overline{\Gamma} t} \sigma^-, \; \cdot \;] ,\\ 
\mathcal{K}^I_2 &=& \mathcal{K}_2.
\end{eqnarray}
~(\ref{eq:NoZero}) satisfies the conditions of Theorem 3 from~\cite{Karasev2023}. Using this Theorem, we can find the correct solution~(\ref{eq:corrsolution}) of~(\ref{eq:limitK2}) (here we apply the limit of large times $t \to t/\lambda^2$, $\lambda \to 0$, $t/\lambda^2 = const$).
\begin{equation}
\frac{d}{dt} \tilde{\rho}(t) = \mathcal{K}_2 (+\infty)\tilde{\rho}(t) \implies \tilde{\rho} (t) = e^{\mathcal{K}_2 (+\infty)t} \mathcal{R} \rho(0),
\label{eq:limitK2}
\end{equation}
\begin{equation}
\mathcal{R} = \mathcal{I} + \lambda \int_{0}^{\infty} dt'\mathcal{K}^I_1 (t').
\end{equation}
Thus renormalized initial conditions are equal to
\begin{equation}
\mathcal{R}\rho(0) = \left( \mathcal{I} - \; i \lambda |\Gamma |^{-2} [gz \overline{\Gamma} \sigma^+ + \overline{g} \overline{z}  \Gamma \sigma^-, \; \cdot \;] \right) \rho(0).
\end{equation}

\section{Quasiprobability distribution}\label{app:qdistibution}

According to~\cite{brewster2018generalized} the corresponding quasiprobability function of the Schrödinger cat state is as follows
\begin{eqnarray}
P(z) &=&|c_\alpha|^2 \Big(\delta^2(z - \alpha) + \delta^2(z + \alpha)
\nonumber \\ 
&&+ e^{i\theta} e^{-2|\alpha^2|} \tilde{\delta}(\Re z + i \Im \alpha) \tilde{\delta}(\Im z - i \Re \alpha)
\nonumber \\
&&+e^{-i\theta}e^{-2|\alpha^2|} \tilde{\delta}(\Re z - i \Im \alpha) \tilde{\delta}(\Im z + i \Re \alpha) \Big),
\nonumber \\
\end{eqnarray}
where $\delta^2(z) = \delta(\Re z)\delta(\Im z)$, and $\tilde{\delta}(z)$ is a so called "Generalized" delta function such that
\begin{equation}
  \int^{\infty}_{-\infty}f(x)\tilde{\delta}(x-z_0)dx = f(z_0)
\end{equation}
So the transformation~(\ref{eq:cattransformation}) is obtained by calculating integrals
$\int P(z)zd^2 z$ and $\int P(z)\overline{z}d^2 z$.

\section{Vectorization}\label{app:vector}
  
It is well-known that the GKSL equation can be formally mapped to the Schrödinger-like equation (generally with non Hermitian Hamiltonian) for the density matrix arranged in a vector \cite{Havel2003}, so it is sometimes called vectorization. We will also use this technique.
  
So we formally map the original Hamiltonians and the dissipator into Hamiltonians with a doubled number of degrees of freedom. Then \eqref{eq:freeGen} and \eqref{eq:intGen} have the vectorized form
\begin{eqnarray}
L_0 &=& - i \omega_A \sigma_l^{+} \sigma_l^{-} + i \omega_A \sigma_r^{+} \sigma_r^{-} - i \omega_c b_l^{\dagger} b_l + i \omega_c b_r^{\dagger} b_r
\nonumber \\
&&+ \gamma\left( b_l b_r - \frac12 b_l^{\dagger} b_l - \frac12 b_r^{\dagger} b_r\right),
\end{eqnarray}
\begin{equation}
L 
= -i g \sigma_l^+ b_l -i \overline{g} \sigma_l^- b_l^{\dagger} + i \overline{g} \sigma_r^+ b_r + i g \sigma_r^- b_r^{\dagger},
\end{equation}
where we have defined $b_l \equiv b \otimes I$ , $b_r \equiv I \otimes b$, $\sigma_l^- \equiv \sigma^- \otimes I$,  $\sigma_r^- \equiv I \otimes \sigma^-$, etc. Application of vectorized creation and annihilation operators to the quadratic GKSL generators was discussed in \cite{prosen2010quantization}.

This way of expanding the initial space is also suitable for more complex models with a different interaction operator, the main thing is that the first part of $L_0$ contains the operators of the finite-dimensional system, and the second part contains quadratic expressions of the creation/annihilation operators.

The dynamics $e^{-L_0 t} \cdot e^{L_0 t} $ will not preserve the self-adjointness of the operators, since we have passed to the effective Hamiltonian. In this case the commutation relations will be preserved at any moment of time. To move into the interaction picture let us define $X(t ) = e^{-L_0 t} X e^{L_0 t} $ for $X \in \{b_l, b_r, b_l^{\dagger}, b_r^{\dagger}\}$. Then the Heisenberg equations with the free generator $L_0$
\begin{equation}
	\frac{d}{dt} X(t) = - [L_0,X(t)]
\end{equation}
take the explicit form
\begin{eqnarray}
\frac{d}{dt}b_l(t) &=& - \left( \frac12 \gamma + i \omega_c\right) b_l(t),\\ 
\frac{d}{dt}b_r(t) &=& - \left( \frac12 \gamma -i \omega_c \right) b_r (t),\\ 
\frac{d}{dt}b_l^{\dagger}(t) &=& \left( \frac12 \gamma + i \omega_c\right) 	b_l^{\dagger}(t) - \gamma b_r(t),\\ 
\frac{d}{dt}b_r^{\dagger}(t) &=& \left( \frac12 \gamma - i \omega_c\right) 	b_r^{\dagger}(t) - \gamma b_l(t).
\end{eqnarray}
This is a closed system of equations that has a simple solution
\begin{eqnarray}
b_l(t) &=& e^{ - \left( \frac12 \gamma + i \omega_c\right) t} b_l,\\
b_r(t) &=& e^{ - \left( \frac12 \gamma - i \omega_c\right) t} b_r,\\
b_l^{\dagger}(t) &=& e^{\left( \frac12 \gamma + i \omega_c\right) t} 	b_l^{\dagger} + e^{ - \left( \frac12 \gamma - i \omega_c\right) t} (1 - e^{\gamma t})b_r,\\
b_r^{\dagger}(t) &=& e^{\left( \frac12 \gamma - i \omega_c\right) t} 	b_r^{\dagger} + e^{ - \left( \frac12 \gamma + i \omega_c\right) t} (1 - e^{\gamma t})b_l.
\end{eqnarray}
Similarly, for two-level operators we have
\begin{eqnarray}
\sigma_l^{\pm}(t) &=& e^{\pm i \omega_A t} \sigma_l^{\pm},\\ 
\sigma_r^{\pm}(t) &=& e^{\mp i \omega_A t} \sigma_r^{\pm}.
\end{eqnarray}
	
It is convenient to move the dependence of $\sigma_l, \sigma_r$ on time to operators $b_l(t)$, i.e. let us represent the interaction Hamiltonian in the interaction picture as
\begin{eqnarray}
L(t) 
&=& -i g \sigma_l^+ B_l(t) -i \overline{g} \sigma_l^- B_l^{\dagger}(t) + i 	\overline{g} \sigma_r^+ B_r(t)
\nonumber \\
&&+ i g \sigma_r^- B_r^{\dagger}(t),
\end{eqnarray}
where we have defined
\begin{eqnarray}
B_l(t) &\equiv& e^{ - \Gamma t} b_l, 
\\ 
B_r(t) &\equiv& e^{ - \overline{\Gamma} t} b_r, 
\\ 
B_l^{\dagger}(t) &\equiv& e^{\Gamma t} 	b_l^{\dagger} + e^{ - \overline{\Gamma} t} (1 - e^{\gamma t})b_r, 
\\ 
B_r^{\dagger}(t) &\equiv& e^{\overline{\Gamma} t} 	b_r^{\dagger} + e^{ - \Gamma t} (1 - e^{\gamma t})b_l. 
\end{eqnarray}
Now we need to introduce a projector consistent with this way of vectorization
\begin{equation}
P = |z\rangle_l  |\overline{z}\rangle_r \langle \mathbbm{1}|.
\end{equation}
On the right side stands a generalized state, such that 
\begin{equation}
\langle \mathbbm{1}| |\psi \rangle_l \overline{| \varphi \rangle}_r = \langle 	\varphi | \psi \rangle.
\end{equation}
For arbitrary operators it follows 
\begin{equation}
\langle \mathbbm{1}| X_l Y_r |\psi \rangle_l \overline{| \varphi \rangle}_r = \langle \varphi | Y^T X| \psi \rangle.
\end{equation}
To calculate the terms of perturbation theory, we will need to find such averages.
\begin{eqnarray}
&&\langle \mathbbm{1}|( b_l^{\dagger})^k b_l^{m} |z\rangle_l |\overline{z}\rangle_r = \langle \mathbbm{1}|( b_l^{\dagger})^k |z\rangle_l |\overline{z}\rangle_r z^{m}
\nonumber \\
&&=\langle z|( b^{\dagger})^k  |z\rangle z^{m} =\overline{z}^k z^{m}, 
\\ 
&&\langle \mathbbm{1}|( b_r^{\dagger})^k b_r^{m} |z\rangle_l |\overline{z}\rangle_r = \langle \mathbbm{1}|( b_r^{\dagger})^k  |z\rangle_l  |\overline{z}\rangle_r \overline{z}^{m}
\nonumber \\
&&= \langle z| b^k  |z\rangle  \overline{z}^{m} = z^k \overline{z}^{m} .
\end{eqnarray}
Further, the main difficulty is to find normal forms from such expressions,
\begin{eqnarray}
&&\langle \mathbbm{1}|( b_l^{\dagger})^{k_l} b_l^{m_l} ( b_r^{\dagger})^{k_r} 	b_r^{m_r}|z\rangle_l  |\overline{z}\rangle_r 
\nonumber \\
&&=\langle \mathbbm{1}|( b_l^{\dagger})^{k_l} ( b_r^{\dagger})^{k_r} |z\rangle_l  |\overline{z}\rangle_r z^{m_l} \overline{z}^{m_r} = \langle z| b^{k_r}( b^{\dagger})^{k_l}  |z\rangle z^{m_l} \overline{z}^{m_r},
\nonumber \\
\end{eqnarray}
which reduces to the representation of $ b^{k_r}( b^{\dagger})^{k_l} $ in the normal order. And for our purposes, it  is appeared to be more convenient to automate this procedure in Wolfram Mathematica. In turn, it allowed us to automate the calculation of $	P e^{L_0 t} L(t) L(t_1)\cdots L(t_n)	P $ and, hence, of \eqref{eq:momentsDef} and cumulants in the vectorized form. 

{
\section{Resonance fluorescence spectrum for depleting field}
\label{app:fluorescence}

The resonance fluorescence spectrum is proportional to the function
\cite[Eqs. (10.5.1)-(10.5.2)]{scully1997quantum}
\begin{equation}
S(\omega) =\operatorname{Re} \int_0^{\infty} \langle \sigma_+(t) \sigma_-(t + \tau)\rangle e^{i\omega \tau}.
\end{equation}

Assuming  that the total dynamics of the atom and the mode of the field is Markovian it can be defined by the  quantum regression formula \cite[Subsec. 3.2.4]{Breuer2002}
\begin{equation}\label{eq:regrCorrFun}
\langle \sigma_+(t )\sigma_-(t+ \tau) \rangle =
\operatorname{Tr}_{S+B}  \mathcal{U}_{\tau}( \sigma_+ \rho_{\rm st}(t) ) \sigma_-, 
\end{equation}
where $\tau > 0$, $t>0$ and $\mathcal{U}_{\tau}$ is the total dynamical map of the atom and the bosonic mode. For the calculation of the Mollow spectrum in the undepleted field,  $ \rho_{\rm st}(t)$ is usually assumed to be in the stationary state and factorized so  such that the quantum regression formula can be applied to the reduced dynamics of the atom \cite[Subsec. 3.4.5]{Breuer2002}. Thus, we use a natural generalization to the case of depleting filed, assuming that in the approximately stationary state is also factorized
\begin{equation}\label{eq:factorAssump}
\rho_{\rm st}(t) \approx (\rho_A)_{\rm st}(t) \otimes |z(t)\rangle\langle z(t)|,
\end{equation}
where $(\rho_A)_{\rm st}(t)$ is an instantaneous stationary state of the generator we are considering, and the field is approximately in the coherent state with $z(t) = z e^{(-i\omega_c - \frac{\gamma}{2}) t }$, consistent with our assumptions for the atomic master equation derivation

Under assumption \eqref{eq:factorAssump} correlation function \eqref{eq:regrCorrFun} takes the form
\begin{equation*}
\langle \sigma_+(t )\sigma_-(t+ \tau) \rangle =
\operatorname{Tr}_{S}  \Phi_{\tau}|_{z(t)}( \sigma_+ (\rho_A)_{\rm st}(t) ) \sigma_-,
\end{equation*}
where $ \Phi_{\tau}|_{z(t))}$ is the dynamical map with generator $\mathcal{K}_{\tau}$ depending on time $\tau$ with  $z(t)$ instead of $z$ as a parameter, so we can apply our perturbative formulae for  $\mathcal{K}_{\tau}$.

}

\bibliography{RefParametric}

\end{document}